%% file: main.tex
\documentclass[10pt,conference]{IEEEtran}

\usepackage[usenames,dvipsnames]{xcolor}
\usepackage{xspace}
\usepackage[hyphens]{url}
\usepackage{hyperref}
\usepackage{cleveref}
\usepackage{booktabs}
\usepackage{multirow}
\usepackage{balance}
\usepackage{microtype}
\usepackage{graphicx}
\usepackage[noadjust]{cite}

\newcounter{insight}
\newcommand{\insight}[1]{\refstepcounter{insight}
	\begin{center}
		\framebox{
			\begin{minipage}{0.93\columnwidth}
				{\bf Insight \arabic{insight}:} \textit{#1}
			\end{minipage}
		}
	\end{center}
}

\makeatletter
\newcommand{\linebreakand}{%
  \end{@IEEEauthorhalign}
  \hfill\mbox{}\par
  \mbox{}\hfill\begin{@IEEEauthorhalign}
}
\makeatother

\begin{document}

\title{Full Line Code Completion: Bringing AI to Desktop}


\author{
\IEEEauthorblockN{Anton Semenkin\IEEEauthorrefmark{2}, Vitaliy Bibaev\IEEEauthorrefmark{2}, Yaroslav Sokolov\IEEEauthorrefmark{2}, Kirill Krylov\IEEEauthorrefmark{2}, Alexey Kalina\IEEEauthorrefmark{2}\IEEEauthorrefmark{1}, Anna Khannanova\IEEEauthorrefmark{2}\IEEEauthorrefmark{1}, \\ Danila Savenkov\IEEEauthorrefmark{2}\IEEEauthorrefmark{1}, Darya Rovdo\IEEEauthorrefmark{2}\IEEEauthorrefmark{1}, Igor Davidenko\IEEEauthorrefmark{2}\IEEEauthorrefmark{1}, Kirill Karnaukhov\IEEEauthorrefmark{2}\IEEEauthorrefmark{1}, Maxim Vakhrushev\IEEEauthorrefmark{2}\IEEEauthorrefmark{1}, \\ Mikhail Kostyukov\IEEEauthorrefmark{2}\IEEEauthorrefmark{1}, Mikhail Podvitskii\IEEEauthorrefmark{2}\IEEEauthorrefmark{1},  Petr Surkov\IEEEauthorrefmark{2}\IEEEauthorrefmark{1}, Yaroslav Golubev\IEEEauthorrefmark{4}, Nikita Povarov\IEEEauthorrefmark{2}, Timofey Bryksin\IEEEauthorrefmark{4}}

\IEEEauthorblockA{\IEEEauthorrefmark{2}\textit{JetBrains}, \IEEEauthorrefmark{4}\textit{JetBrains Research} \\
\{anton.semenkin, vitaliy.bibaev, yaroslav.sokolov, kirill.krylov, alexey.kalina, anna.khannanova, \\ danila.savenkov, darya.rovdo, igor.davidenko, kirill.karnaukhov, maxim.vakhrushev, \\ mikhail.kostyukov, mikhail.podvitskii, petr.surkov, yaroslav.golubev, nikita.povarov, timofey.bryksin\}@jetbrains.com}

}

\maketitle

\begin{abstract}
  In recent years, several industrial solutions for the problem of multi-token code completion appeared, each making a great advance in the area but mostly focusing on cloud-based runtime and avoiding working on the end user's device.
  
  In this work, we describe our approach for building a multi-token code completion feature for the JetBrains' IntelliJ Platform, which we call \textit{Full Line Code Completion}. 
  The feature suggests only syntactically correct code and works fully locally, \textit{i.e.}, data querying and the generation of suggestions happens on the end user's machine.
  We share important time and memory-consumption restrictions, as well as design principles that a code completion engine should satisfy. 
  Working entirely on the end user's device, our code completion engine enriches user experience while being not only fast and compact but also secure.
  We share a number of useful techniques to meet the stated development constraints and also describe offline and online evaluation pipelines that allowed us to make better decisions.
  
  Our online evaluation shows that the usage of the tool leads to 1.3 times more Python code in the IDE being produced by code completion. 
  The described solution was initially started with a help of researchers and was then bundled into all JetBrains IDEs where it is now used by millions of users. Thus, we believe that this work is useful for bridging academia and industry, providing researchers with the knowledge of what happens when complex research-based solutions are integrated into real products.
\end{abstract}


\input{sections/01-intro}
\input{sections/02-related-work}
\input{sections/03-approach}
\input{sections/04-evaluation}
\input{sections/05-discussion}
\input{sections/06-threats}
\input{sections/07-conclusion}

\bibliographystyle{IEEETran}
\balance
\bibliography{IEEEabrv,literature}

\end{document}

%% file: sections/01-intro.tex
\section{Introduction}

\renewcommand*{\thefootnote}{\fnsymbol{footnote}}
\footnotetext[1]{\vspace{-0.2cm}These authors contributed equally to the work.}
\renewcommand*{\thefootnote}{\arabic{footnote}}

Code completion is among the most attractive features in integrated development environments (IDEs)~\cite{eclipse-usage-2006, vs-usage-2016}, since it helps developers to avoid wasting time on manually typing all of the code characters and thus speeds up any software engineering activity.
Depending on the specific IDE, the complexity of code completion (and thus its quality) varies a lot: it may just complete strings that a developer has already used in the current file, or it can use more complex approaches like static code analysis to predict possible tokens that a programmer may currently need.
In any case, built-in completion engines usually only work in a single-token mode, \textit{i.e.}, only one identifier is generated (\textit{e.g.}, a variable or a method name). 

In the past years, several products for \textit{multi-token} code completion have appeared on the market, including Copilot~\cite{copilot}, TabNine~\cite{tabnine}, Codeium~\cite{codeium}, Sourcegraph Cody~\cite{cody},  aiXcoder~\cite{aixcoder}, and others.
Most of such tools are cloud-based, so they miss a wide niche of customers, which includes users beyond a firewall, as well as users with limited internet access.
While TabNine and aiXcoder support fully local models and data querying, they lack integration features and do not implement code correctness checks.
This still leaves an open market for code completion engines that are local-based and ensure a proper user experience in an IDE. 

In recent years, machine learning techniques were studied and applied for the different aspects of the completion problem~\cite{codex-paper, bibaev2022all, whole-line, svyatkovskiy2021fast}, including the generation of the completion sequences itself.
Since source code can be viewed as sequential data, natural language processing (NLP) methods have found their application for the task~\cite{software-naturalness-2016}.
Specifically, Transformer-based~\cite{attention-is-all} neural networks significantly outperform other architectures in generating sequential data~\cite{gpt2, gpt3, t5, Hellendoorn2020Global}.

At JetBrains, we built the \textit{Full Line Code Completion (FLCC)} feature~\cite{full-line} --- a multi-token code completion engine that utilizes a Transformer-based neural network in its core and suggests only syntactically correct code.
The feature is implemented in the form of a plugin for the IntelliJ Platform~\cite{intellij, kurbatova2021intellij} (an open-source platform for building IDEs), and is bundled into all JetBrains IDEs: IntelliJ IDEA~\cite{ij-idea}, PyCharm Pro~\cite{pycharm}, WebStorm~\cite{webstorm}, and others, where it is enabled by default for various programming languages.
The feature has been available for Python since version 2023.3 and then became available for other languages too in 2024.1 and 2024.2, now serving millions of users.
It works fully locally, \textit{i.e.} all computations, from code completion invocation to the showing of suggestion, happen on the end user's device.
Following modern industrial standards, the plugin is implemented in Kotlin, however, it utilizes an additional native server that is run locally and is implemented in C++.
The whole model training pipeline is written mostly in Python, reusing the Kotlin pipeline only for the data processing step.

Even though systems employing Transformer-based models comprise state-of-the-art solutions for the generation of multi-token sequences, they typically employ infrastructure that involves cloud computations: models have gigabytes of memory footprint and demand high-performance computational devices like GPUs.
This created a general challenge for us: bringing such functionality to the end user's local device takes both engineering effort and the utilization of many hacks and tricks that we are going to share in this paper.
Moreover, the existing design of IntelliJ IDEs focused on standard one-token code completion in the form of a lookup window with code suggestions (see Figure~\ref{fig:completion}a).
This brings an additional development challenge --- Full Line Code Completion should be \textit{properly integrated} into an IDE, meaning that it should smoothly enhance the user's experience rather than break the existing habits and usage patterns.
In summary, we formulated and fulfilled the following restrictions and design principles for our Full Line Code Completion feature:

\begin{enumerate}

    \item The model and its execution engine must be \textit{local-based}, \textit{i.e.}, operate fully on the user's machine without sending anything over the internet.

    \item Solution's \textit{memory footprint must be feasible} and take a reasonable part of the IDE's memory footprint. 

    \item The solution must be \textit{fast}, \textit{i.e.} it must feel seamless to the developer while typing.

    \item The feature must be \textit{properly integrated} into the existing code completion usage pipeline of developers. 
\end{enumerate}

In this paper, we describe how we built this Full Line Code Completion feature.
Starting from the study of existing datasets, data collection, and the pre-processing with multiple tokenization specifics, we describe the details of the neural network training process and the end-user runtime, including a modified beam search algorithm, post-processing features, and user experience (UX) choices.
While we cannot provide a full replication package of the described product feature, we share multiple conceptual insights we derived from our experience.
The focus of the paper is on the Full Line Code Completion for Python, for which 100M parameters LLaMA-like~\cite{llama} model was used, and inline completion (\textit{i.e.}, gray text completion) mode was implemented as a part of the IntelliJ Platform (see Figure~\ref{fig:completion}b).
Despite mainly focusing on Python, we share some evaluation results for other languages as well to provide a fuller picture.
While preparing for the releases, we conducted rigorous offline evaluation, as well as online evaluation in the form of A/B testing. 
Our experiments showed that the ratio of completed code (\textit{i.e.}, the percentage of code symbols written with code completion among all written code) increased 1.3 times for Python users with Full Line Code Completion compared to users with standard code completion only.

Overall, our contributions are the following:
\begin{itemize}
    \item We provide an in-depth exploration of the implementation of Full Line Code Completion for 2024.1 \& 2024.2 releases of IntelliJ-based IDEs.
    \item We share details of offline and online evaluation setups, as well as the derived experimental results, which show significant benefits of using Full Line Code Completion.
    \item We list a set of specific problems we encountered while implementing Full Line Code Completion and our solutions for them. 
    We believe that this information can serve for bridging academia and industry, providing researchers with the knowledge of what happens when complex research-based solutions are integrated into real products.
\end{itemize}

The rest of the paper is organized as follows. 
In Section~\ref{sec:bg_and_rw}, we describe the related work in the field. 
Then, in Section~\ref{sec:approach}, we describe our approach in detail, highlighting various problems we encountered and how we solved them. 
Section~\ref{sec:evaluation} presents the results of our A/B testing experiments, while Section~\ref{sec:discussion} discusses these results and our lessons learned. 
Finally, in Section~\ref{sec:ttv}, we describe the threats to the validity of our work, and in Section~\ref{sec:conclusion}, we conclude the paper.

%% file: sections/02-related-work.tex
\section{Background and Related Work}
\label{sec:bg_and_rw}

\subsection{Background}
\label{sec:bg_and_rw:bg}

\subsubsection{Code completion}
\label{sec:bg_and_rw:bg:code_completion}
\begin{figure}[t]
  \centering
  \includegraphics[width=\columnwidth]{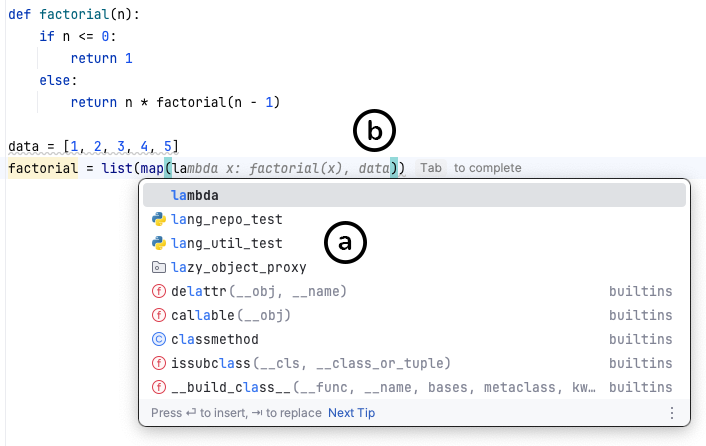}
  \caption{Code completion in the IDE: \textbf{(a)} standard lookup window, and \textbf{(b)} multi-token inline (gray text) completion.}
  \label{fig:completion}
  \vspace{-0.5cm}
\end{figure}

Code completion, a prevalent feature across various IDEs~\cite{eclipse-usage-2006, vs-usage-2016}, is aimed at optimizing the work of a developer, not only by speeding up code typing but also by facilitating API exploration.
Common in modern IDEs, \textit{single-token}, or \textit{standard}, code completion suggests individual tokens via a lookup window (see Figure~\ref{fig:completion}a). 
This feature, relying on static code analysis, generates suggestions from multiple \textit{candidate providers}, which include both built-in IDE modules and external plugins.

In addition to recent advancements in single-token code completion, there is a niche of \textit{multi-token} completion that is being actively studied in academia~\cite{codegen-paper, codellama-paper, incoder-paper, starcoder-paper} and employed in products~\cite{copilot, codewhisperer, tabnine, codeium}.
Similar to single-token completion, multi-token completion can be obtained from a corresponding candidate provider.
However, in contrast to standard code completion, it is usually shown as inline completion (also referred to as \textit{gray text completion}, see Figure~\ref{fig:completion}b) and thus usually provides only one suggestion.

Most multi-token code completion solutions in both academia and industry tend to use language modelling approaches to generate suggestions.
Specifically, they utilize Transformer-based neural networks~\cite{attention-is-all} like GPT-2~\cite{gpt2} and LLaMA~\cite{llama}.
Such models usually have billions of parameters, implying gigabytes of memory footprint. 
Moreover, they are computationally intensive, so cloud-based multi-GPU servers are required to provide candidate suggestions.

\subsubsection{Language modelling}
\label{sec:bg_and_rw:bg:language_modelling}
Over the past few years, Transformer-based neural networks started dominating in language modelling~\cite{gpt2, gpt3, llama}.
Specifically, models based solely on the decoder part of a Transformer (also known as \textit{autoregressive} Transformers) are utilized for multi-token generation.
A typical multi-token generation pipeline includes tokenization and sequence generation algorithms.
Below, we briefly describe these parts for clearer understanding of the challenges we encountered when implementing FLCC. 

\textbf{Tokenization.} 
There exist many techniques for breaking down source sequences into tokens, including character-level, word-level, and subword-level tokenizations. 
The latter is typically used~\cite{attention-is-all, bpe-nlp} with a Transformer-based model of any kind.
Specifically, byte-pair encoding (BPE)~\cite{gpt2, llama} is used for decoder-based autoregressive Transformers like GPTs and LLaMAs.
This algorithm iteratively combines the most frequent pairs of adjacent tokens (bytes, characters, or character sequences) to form new, longer tokens. 
This process is repeated a predefined number of times or until a desired vocabulary size is reached.
Then, during tokenization, the source sequence is transformed into a list of integers that correspond to an ordinal number of each token is the combined vocabulary.

\textbf{Sequence generation algorithm.}
Transformer decoders compute pair-wise relations between all tokens in the given context to produce a probability distribution over the next token.
To obtain multiple tokens using such a model, one needs to apply it several times in a row, with each next distribution depending on the token generated on the previous step. 

This raises the problem of the search for the most probable token sequence. 
Since BPE vocabulary usually consists of thousands of tokens, the generation of the next step distribution for every possible next token is not feasible.
One of the popular solutions to this problem is the \textit{beam search} algorithm~\cite{hf_generation}.

\begin{figure}[b]
  \centering
      \vspace{-0.3cm}
  \includegraphics[width=\columnwidth]{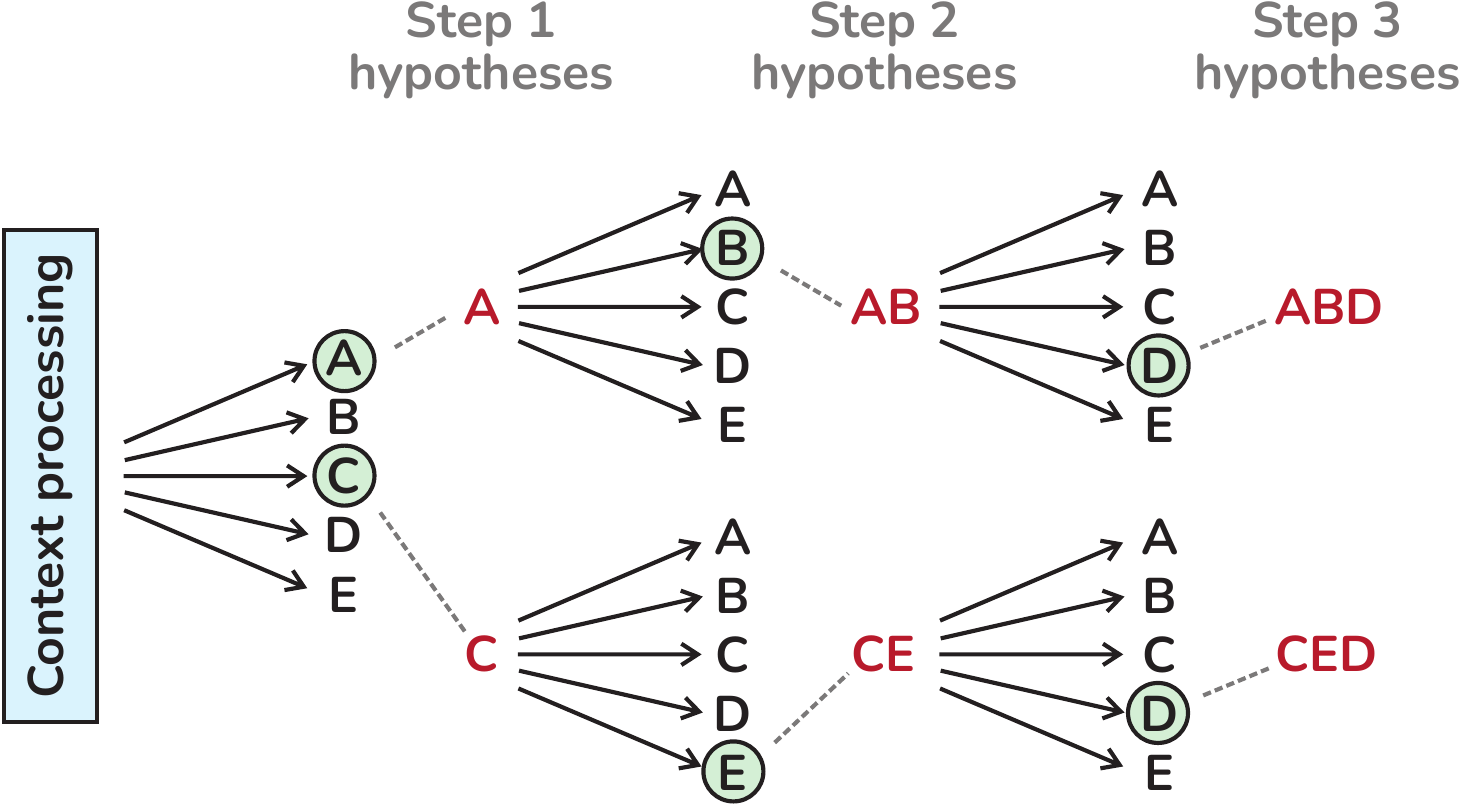}
  \caption{The general pipeline of a beam search algorithm.}
  \label{fig:beam_search}
\end{figure}

The overall scheme of the algorithm is presented in Figure~\ref{fig:beam_search}. 
Beam search is a heuristic algorithm that focuses on traversing the graph of all possible sequences by expanding the node with the highest potential. 
This approach maintains only a fixed number of potential solutions at each step, called \textit{beams}, or \textit{hypotheses}. 
By focusing on the most promising hypotheses and discarding the less likely ones, beam search efficiently navigates through the graph.
Usually, top $k$ most probable beams are supported at each step, and the next hypothesis score is obtained from the previous one, multiplying it by the probability of the next token. 
For Transformers, this procedure has an efficient implementation: the relations between context tokens can be computed once and reused afterward.

\subsubsection{Release cycle of IDEs at JetBrains}
\label{sec:bg_and_rw:bg:intellij_releases}
At JetBrains, an established IDE release cycle comprises 3 major releases per year, enumerated by year and year's version in the following format: $YYYY.X$.
Each release is preceded by a free early access program (EAP), allowing users to download experimental versions of the IDEs. 
Participation in the EAP includes user consent for the anonymized submission of usage logs to JetBrains.
This data is crucial for our analysis, enabling us to determine the users' experiences with new features.
EAP releases are also used as a platform for A/B testing.
The EAPs of versions 2023.3, 2024.1, and 2024.2 were used to conduct the online evaluation for Full Line Code Completion.

\subsection{Related Work}
\label{sec:bg_and_rw:rw}

\subsubsection{Datasets}
\label{sec:bg_and_rw:rw:datasets}
Many existing datasets for training code completion models, particularly for Python~\cite{codexglue, slm-data, 150k-python}, have limitations due to licensing issues and quality concerns. 
Open-source code, often collected from platforms like GitHub, may have restrictive licenses or none at all~\cite{golubev2020study}, which poses legal challenges~\cite{licenses-lawsuit}, as seen in GitHub Copilot's~\cite{copilot} usage. 
Furthermore, direct extraction of GitHub's Python files yields many autogenerated and duplicated files~\cite{the-stack-paper}, leading to potential biases and data leakage issues in model training. 
A notable development to address these issues is the introduction of The Stack dataset~\cite{the-stack, codex-paper, the-stack-paper} by BigCode~\cite{bigcode}, a collaboration of over 350 researchers backed by HuggingFace and ServiceNow. 
This dataset, encompassing over 6TB of code in 358 languages, is extensively used for training large language models due to its permissive licensing and wide coverage.

\begin{figure*}[t]
  \centering
  \includegraphics[width=\textwidth]{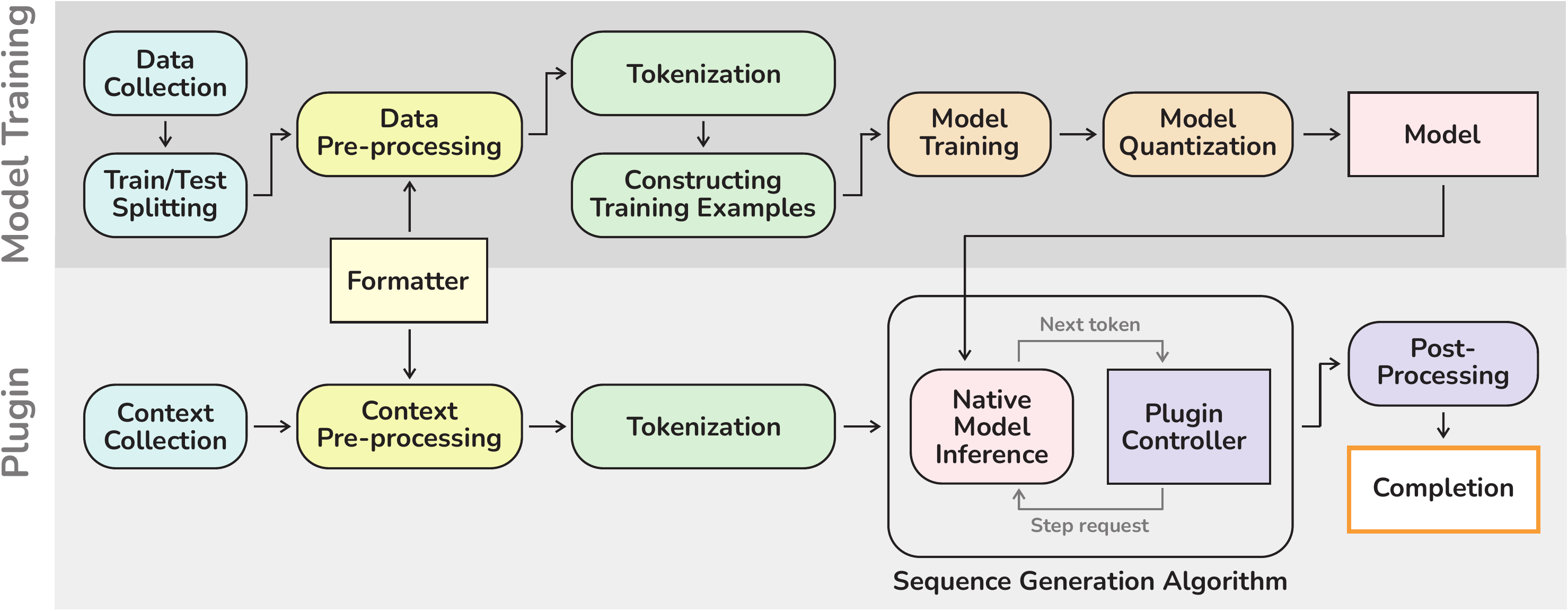}
  \caption{The overall pipeline of Full Line Code Completion.}
  \label{fig:pipeline}
  \vspace{-0.2cm}
\end{figure*}

\subsubsection{Code completion models}
\label{sec:bg_and_rw:rw:completion_models}
In recent years, code completion became a major field of study for deep learning applications.
Many researchers and companies trained and released open-source models, resulting in the availability of training code, configuration specifics, and models' weights. 

The employed models are typically decoders of the Transformer architecture, including GPT and LLaMA architectures representing specific types of decoders. 
The most famous examples of such models are InCoder~\cite{incoder-paper, incoder-repo}, StarCoder~\cite{starcoder-paper, starcoder-repo}, CodeLLaMA~\cite{codellama-paper, codellama-repo}, CodeGen and CodeGen-2~\cite{codegen-paper, codegen2-paper, codegen-repo}, CodeParrot~\cite{codeparrot-repo}, and PolyCoder~\cite{polycoder-paper, polycoder-repo}.
Despite the fact that the models are open-source, they are not suitable for local code completion engines like FLCC for several reasons.
First of all, they are usually too big to ship to users and do not meet the memory footprint restrictions for desktop applications like IDEs.
Secondly, the license for the model's weights might not be permissive for commercial use.
Last but not least, only specific architectures like GPT-2 and LLaMA have corresponding inference engines that allow to run them on an arbitrary end-user device fast enough.

\subsubsection{Multi-token code completion products}
\label{sec:bg_and_rw:rw:completion_products}
Lately, several products for multi-token completion appeared on the market, including Copilot~\cite{copilot}, TabNine~\cite{tabnine}, Codeuim~\cite{codeium}, CodeWhisperer~\cite{codewhisperer}, aiXcoder~\cite{aixcoder}, Sourcegraph Cody~\cite{cody}, and others.
Some of them remain free to use (\textit{e.g.}, Codeium and CodeWhisperer), while others require paid subscription.
More importantly, all such products, except for TabNine and aiXcoder, offer only cloud-based solutions that might have certain restrictions for users behind the firewall, users with limited internet access, and users who have privacy concerns regarding sending their code to the cloud.
Cloud-based solutions are unavailable not only for huge markets like China, but also for enterprise clients that want the source code to stay inside their network.
Moreover, none of these solutions have a common API for inline completion, which makes them conflict with others and requires re-implementing the same functionality, which leads to poor UX and many confirmed bugs.

%% file: sections/03-approach.tex
\section{Approach}
\label{sec:approach}

Considering everything described in the previous section, our goal was to create a Full Line Code Completion IDE feature that adheres to four important restrictions: it should be local-based, compact, fast, and well-integrated. 

\begin{enumerate}

    \item The model and its execution engine must be \textit{local-based}, \textit{i.e.}, operate fully on the user's machine without sending anything over the internet. 
    This is crucial because of privacy concerns and possible internet connection issues.

    \item Solution's \textit{memory footprint must be feasible} and take a reasonable part of the IDE memory footprint. 
    This ensures a convenient installation and a feasible runtime, not overloading the users' machines for a single functionality.
    Insights from the IDE development teams give us a general upper bound of about 20--40\% of the IDE memory footprint (might differ between products).

    \item The solution must be \textit{fast}, \textit{i.e.} it must feel seamless for developer while typing.

    \item The feature must be \textit{properly integrated} into the existing code completion usage pipeline of developers. 
    FLCC must be non-intrusive, enhancing the coding process by offering helpful suggestions, without introducing wrong or irrelevant code.
    Additionally, in IntelliJ-based IDEs, multiple \textit{providers} might contribute both to the standard lookup completion window and to inline completion.
    So, being a default functionality, FLCC must not provoke usage conflicts with either standard or third-party code completion providers.
\end{enumerate}

The general pipeline of our approach is shown in Figure~\ref{fig:pipeline}. 
It consists of two main parts: the model training pipeline (top) and the code completion runtime that is executed on the end user's device inside the plugin (bottom).
The first segment can be decomposed into collecting the data, pre-processing it, tokenizing it, and training a Transformer-based model on it.
The second segment encompasses the same pre-processing used during training, as well as the model's runtime, post-processing features, and inline completion API used to show the suggestion and interact with the user.
The training part is developed in Python, whereas the second part predominantly uses Kotlin, except for the inference process, implemented in C++.
The training pipeline also incorporates code for some of the data pre-processing steps from the plugin.
Now, let us describe each step in greater detail.

\subsection{Data Collection}
\label{sec:approach:data_collection}

\subsubsection{Dataset}
\label{sec:approach:data_collection:dataset}
For Python, we used the subset of The Stack~\cite{the-stack} --- the StarCoder dataset~\cite{starcoder-data}, which we additionally processed to keep only permissive licenses.
This completely fulfilled our needs in terms of handling licenses, as well as the filtering of duplicated files.

\subsubsection{Train/test splitting}
\label{sec:approach:data_collection:train_test}
For training a model, the data is divided into training, validation, and testing segments.
The training set is used for the primary adjustment of the model. 
The validation set allows for intermittent evaluation of the model as it is being trained. 
After training, the testing set evaluates the model's performance.

After collecting the dataset, we found a representative set of repositories to be separated into the test part, thus ensuring that the model is not trained on them. 
This data is used later in a specialized offline evaluation setup described in Section~\ref{sec:evaluation:offline}.
Additionally, to avoid potential data leaks, we ensure that all the forks of the given repository are placed in the same data segment. 
For the training of the Python model, the corresponding proportions of the train-validation-test sets of the data were 80\%-5\%-15\%, adding up to slightly more than 50 GB of data.

\subsection{Data Pre-processing}
\label{sec:approach:preprocessing}
After the initial data collection, it is important to standardize training examples and make them cleaner so the model can learn only useful information and avoid redundant noise. 
For each language, and for Python in particular, we implement this procedure in the form of \textit{Formatter} --- an IntelliJ Platform module that can be used externally from the model training pipeline and that is used on the end user's device while performing the actual code completion. 
Formatter implements the steps we describe below.

\textbf{Removing comments.} Since we are focused on code completion, we filtered out all comments.
Because our solution is limited in terms of model size, it is especially crucial to focus on a single data domain and learn the structure of the source code itself rather than the natural language that describes it.
Additionally, we removed empty lines and trailing whitespaces since they do not contribute to source code logic.

\textbf{Transformation of the indentations.} It is also useful to remove leading spaces and tabs from every line so that the total amount of characters in the training example lowers.
However, for indentation-sensitive languages like Python, these whitespaces are crucial parts of the source code because they indicate the scope of the line. 
We mitigated this by removing all leading indentations and substituting them with special tokens \texttt{<SCOPE\_IN>} and \texttt{<SCOPE\_OUT>}.
While training the tokenizer, we forbid merging such tokens with other tokens so that the acquired vocabulary does not contain any tokens that differ only in leading whitespaces.

Without this step, one will end up with a BPE vocabulary consisting of lots of similar tokens that differ only in the number of tabs in the beginning. 
The examples are: 

\begin{itemize}
    \item \texttt{for i in range(}
    \item \texttt{$\backslash$t for i in range(}
    \item \texttt{$\backslash$t $\backslash$t for i in range(}
\end{itemize}

Such an approach allows not only to save space in the BPE vocabulary but also to better handle the token healing problem that is described further in Section~\ref{sec:approach:sequence:token_healing}.

\insight{Repeating tabulation on different indentation levels can result in a lot of similar tokens in the vocabulary, which might lead to various inefficiencies for the model. 
To combat this, we introduce two special tokens that denote the change of scope.}

\textbf{Import dropout.}
The final step in preprocessing the data for training involves adapting to the workflow in IDEs, in particular, to how developers handle code and imports.
Developers often write code first and add imports later, a process facilitated by the \textit{auto-import} feature in IntelliJ-based IDEs --- it automatically recognizes code from external libraries and adds the necessary imports. 
During training, it is crucial to mimic this behavior by teaching the model to predict tokens from libraries even before they are imported. 
To achieve this, we randomly remove each import statement from training files with a 50\% probability, helping the model to learn to anticipate the necessary imports without their explicit inclusion.

\insight{To account for the auto-import feature and to mimic it in training, we can randomly remove some imports and train a model to suggest tokens without the import in the context.}

We used to have an additional step while prototyping FLCC --- formatting all the code (both on training and inference steps) with the \textit{Black} formatter~\cite{black}, which is a popular strict formatter for Python.
On the one hand, such procedure allows to simplify stylistic patterns in the code, so it is easier to train the model towards better quality. 
On the other hand, we have discovered that such strict pre-processing leads to a massive problem where generated suggestions do not match the code style of the user's project.
Moreover, such formatting requires code in the end user's file to be fully correct, which is not always the case when invoking code completion.
Finally, having such a formatting feature requires reimplementing Black in Kotlin so it can be shipped to the end user as a part of FLCC functionality.
Thus, we removed this step.

\insight{Strict code formatting like Black for Python might be beneficial when validating on the formatted data, however, in real-world applications, it leads to code suggestions mismatching the user's code style or might be inapplicable at all.}

As a result, we managed to implement language-specific Formatter logic in the FLCC plugin in Kotlin that supports a minimal number of straightforward features and can be used as a module from the model training pipeline written in Python.

\subsection{Model Training}
\label{sec:approach:model_training}
The main part of the Full Line Code Completion is the model that is used to generate code suggestions. 
In our approach, the way to obtain the final model has several steps, including tokenization, constructing training examples, training the Transformer-based model, and then quantizing it.

\subsubsection{Tokenization}
\label{sec:approach:model_training:tokenization}
In FLCC, we use Transformer-based models for the source code, so we chose to use the byte-pair encoding (BPE)~\cite{bpe-nlp} as a standard approach for such models~\cite{gpt2, gpt3, llama}. 

\textbf{Character-pair encoding.}
The traditional BPE is supposed to be used for bytes directly, so it is applicable to all languages if we choose an encoding like UTF-8. 
However, we decided to use \textit{character}-pair encoding with slight modifications. 
Compared to natural language texts, source code primarily consists of common statements or sequences of more than one token~\cite{shin2019program}. 
Some good examples of such recurring ``idioms'' in Python are the following:
\begin{itemize}
    \item \texttt{for i in range(}
    \item \texttt{return True}
    \item \texttt{if \_\_name\_\_ == "\_\_main\_\_":}
\end{itemize}

Thus, in our approach, we implement character-pair encoding on top of the YouTokenToMe (YTTM) library~\cite{yttm}, allowing to merge pairs over tabs and spaces but forbidding merges over the newline (\texttt{\textbackslash n}).
This approach allows us to create better representations of the lines in the source code and compress it more efficiently. 
This becomes significantly important when generating full lines of code because a full line then requires fewer tokens, hence leading to less computational time spent on executing the model. 
A similar approach is used by Meta Research for the Incoder model~\cite{incoder-paper}.

\insight{To compress the code better, we can augment the default byte-pair encoding to be character-pair, as well as allow the merging of tokens over the spaces and tabs. 
This way, recurring idioms become single tokens and can be generated more easily.}

We chose the vocabulary size of 16,384 tokens because it allows for a decent model quality while maintaining an acceptable memory footprint: for our sizes of models, a significant part of the model's weights depends on the vocabulary size.

\textbf{Cleaning vocabulary.}
After the initial experiments with tokenization, we observed an important peculiarity of the obtained tokens that had to be dealt with.
For a vocabulary of 16,384 tokens, as many as 800 tokens consisted of characters from various languages, mostly Chinese, but also Japanese, Korean, Arabic, and others. 
Since there are no comments in the dataset, these characters came from string literals and are not suited for code generation.
This observation led us to force only ASCII symbols to be used for the vocabulary while leaving other symbols as \texttt{<UNK>} tokens and merging sequences of them into a single \texttt{<UNK>} token.

\insight{Even without the comments, raw code also contains characters that can hinder the model's ability to suggest proper tokens. 
To combat this, we focused on the ASCII symbols.}

\subsubsection{Constructing training examples}
\label{sec:approach:model_training:example}
We construct each training example as follows. 
Firstly, we sample a file from the train set.
Then, we concatenate file extension, \texttt{<LANG\_SEP\_CHAR>}, file path, \texttt{<METAINFO\_SEP\_CHAR>}, and the part of the code above caret, such that the total amount of tokens in the combined three parts fits the maximum context size of the model.
This way, the meta information is always present in the context given to the model, like a header.

\subsubsection{Base model}
\label{sec:approach:model_training:base}

Having experimented with several model architectures, including GPT-2 and LLaMA, we ended up with a LLaMA-like model for FLCC for all languages. 
In our evaluations, LLaMA-like models performed best not only in terms of quality but also speed, thanks to llama.cpp~\cite{llama-cpp} --- a~multiplatform inference engine for LLaMA-like models.

We use the HuggingFace~\cite{transformers-hf} implementations to train models on the described dataset. 
The models are trained in an automatic mixed precision mode~\cite{mixed_precision} for several days on 8 NVidia H100 GPUs. 
Following recent advancements in choosing the number of training steps given the model and dataset sizes, we used HuggingFace's approximation~\cite{hf-optimal} for every trained model.
For our models, we used the maximum context length of 1,536 tokens, since it performed the best according to our benchmarks, given memory footprint restrictions from target IDEs.
Unlike in many similar circumstances, we do not employ knowledge distillation~\cite{distilbert}, because we discovered that for our sizes of the models, this technique does not provide any quality improvement.

To control the model's overfitting while training, we use a standard cross-entropy loss function evaluated on the validation dataset. 
However, we pay little attention to well-known metrics like Recall at k \textbf{(R@k)} or Mean Reciprocal Rank at k \textbf{(MRR@k)} because they evaluate only single-token prediction quality, whereas FLCC generates multi-token suggestions.

\subsubsection{Quantization}
\label{sec:approach:model_training:quantization}
There are memory footprint and speed constraints among restrictions for FLCC.
Thus, to compress the model's size and speed up its inference, a well-known technique called \textit{quantization}~\cite{quantization-survey} was used.
Quantization refers to changing the way the model weights are stored and changing the precision that is used for model-related computations. 
For LLaMA-like models, quantization from FP32 (32-bit floating point numbers) to INT4 (4-bit integer numbers) is supported via llama.cpp.
This transformation decreases the size of the model several times, trading it off to a negligible quality loss.

The quantization allowed us to achieve two major goals. 
Firstly, the size of the model decreased from almost 400 MB to slightly more than 100 MB, which fits within the desired restrictions. 
Secondly, llama.cpp allows us to shorten the CPU inference time due to more efficient usage of the registers for INT4 data: \textit{e.g.}, the quantized model runs 2 times faster compared to full precision model on M2 CPU on MacOS and 4 times faster on Intel i9 CPU on Linux.

In summary, for Full Line Code Completion, INT4-quantized LLaMA-like models with 100M parameters were used.
The models utilize a modified BPE tokenizer implemented on top of the YouTokenToMe library.

\subsection{Sequence Generation Algorithm}
\label{sec:approach:sequence}
Transformer-based models like GPT-2 and LLaMA are trained to predict the probability distribution over the vocabulary tokens given a context. 
Simpler put, such models predict which token should be next, given some context: in our case, file path, file extension, and lines of code above the caret.
Thus, after the model is trained, it is necessary to wrap it in the \textit{sequence generation algorithm} so that the sequence of several tokens can be generated.
For Full Line Code Completion, we implemented beam search with several key modifications compared to the default approach. 

\subsubsection{Token healing}
\label{sec:approach:sequence:token_healing}
\begin{figure}[t]
  \centering
  \includegraphics[width=\columnwidth]{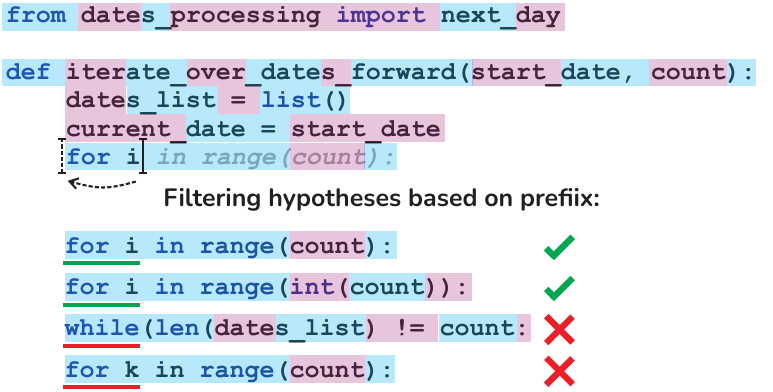}
  \caption{\textit{Token healing}. When the caret falls into the middle of a BPE token, we go left until its start (top) and then filter out hypotheses that do not start with the prefix (bottom).}
  \label{fig:token_healing}
\end{figure}

In real-world code completion, the caret position can arbitrarily divide a token, leading to tokenization being inconsistent with the model's training. 
An example case is presented in Figure~\ref{fig:token_healing}.
Here, code completion is invoked after ``\texttt{for i}'', when the programmer intended to write ``\texttt{for i in range(count):}'', and this creates a tokenization issue. 
During training, the model sees this as a single ``\texttt{for i in range(}'' token. 
However, in this context, this might be split into ``\texttt{for }'' and ``\texttt{i}'', differing from the training scenario and reducing the generation quality. 
This problem escalates with larger average token lengths, increasing the likelihood of mid-token caret positions.

To address this issue in FLCC, we find an offset in the line, such that tokenization of the code to the left of the offset resembles training tokenization more. 
This involves backtracking left from the caret position for as long as the slice from the offset to the caret position is a full substring of one of the vocabulary tokens.
Then, during model execution, we can filter out hypotheses that do not match the cut-out characters.

For example, in Figure~\ref{fig:token_healing}, starting from ``\texttt{for i}'' and moving left until the start of the line, the prefix matches the vocabulary token ``\texttt{for i in range(}''. 
During code suggestion, the process begins with an empty prefix, discarding hypotheses that do not align with the ``\texttt{for i}'' prefix. 
This method, called \textit{token healing}, resolves generation issues from mid-token positions by reverting to one token prior and constraining the initial generation to the incomplete token.

\insight{Having longer tokens allows us to suggest full lines more conveniently, however, this also introduces a problem of generation starting from mid-token position, which can be solved by the token healing technique.}

\subsubsection{Collecting hypotheses}
\label{sec:approach:sequence:stashing}
During beam search, one obtains a set of the most probable hypotheses that continue the context initially fed to the model. 
In our experiments, we found that collecting only those hypotheses that end with \texttt{\textbackslash n} (even if they are not the most probable ones) is the most beneficial approach. 
We call such hypotheses \textit{terminated}.

The rationale beyond this change is the following.
The sequences that were not finished with \texttt{\textbackslash n} in the maximum allowed number of steps in the sequence generation algorithm are probably too long and might not finish at all even if the algorithm continued working.
Moreover, the user expects a code-suggesting tool to show finished constructions.

\insight{Collecting end-of-line-finished hypotheses allows suggesting better completion candidates.}

\subsubsection{Dynamic number of beam search iterations}
\label{sec:approach:sequence:dynamic}
Another modification to the standard algorithm that we implemented is that the number of iterations we perform in the beam search is not pre-defined but dynamic.
More precisely, the generation algorithm has the following set of stopping criteria:
\begin{itemize}
    \item The number of iterations reached the maximum allowed number $n$.
    \item All current hypotheses are terminated.
    \item All current non-terminated hypotheses have $k$ times lower probability than the best terminated hypothesis.
\end{itemize}

The rationale for this is the following. 
FLCC is expected to generate sequences until the end of the line, however, a plain increase in the number of iterations might lead to higher computational costs. 
Thus, we need to continue generating or collecting hypotheses only in scenarios that look promising enough.
Our experiments showed that $n=20$ and $k=3$ are the best values in our setup.

\insight{A dynamic approach in beam search with carefully selected stopping criteria allows generating longer suggestions while keeping the mean computation time almost the same.}

The sequence generation algorithm is used both for studying preliminary metrics during training and inside the FLCC plugin.

\subsection{Plugin}
\label{sec:approach:plugin}
The Full Line Code Completion plugin for the IntelliJ Platform is the place where the described code completion happens. 
It implements all the necessary steps: starting from the invocation of code completion and model inference and finishing with the post-processing and showing of the suggestion.
The model training pipeline is implemented in Python, so it is not possible to directly reuse code for the plugin, which is written in Kotlin.
Thus, we ensure that the processes required for both stages are implemented to be as similar to each other as possible~\cite{rules_ml}.
This includes:
\begin{itemize}
    \item \textbf{Data processing}. In the plugin, the code from the currently opened file in the editor is taken along with its path and extension. 
    The same \textit{Formatter} code is applied to it, and the context for the model is composed in the same manner as described in Section~\ref{sec:approach:preprocessing}.
    \item \textbf{Sequence generation algorithm}. The algorithm described in Section~\ref{sec:approach:sequence} is implemented in Kotlin for the plugin. 
\end{itemize}

When transferring the prototype from research into the product, it is important to remember that the quality of the feature is predicated not only on the model's quality but also on various plugin-related features and UX decisions.
This section describes our engineering efforts comprising FLCC's high quality and speed.

\subsubsection{Inference}
\label{sec:approach:plugin:inference}
Full Line Code Completion is a computationally intensive plugin, because it involves the execution of a neural network.
To achieve the best possible efficiency and speed, we employed the latest developments of inference engines for Transformer-based models --- native CPU runtime for model inference. 
The implementation is as follows:
\begin{itemize}
    \item Alongside FLCC in the running IDE, a separate instance of a native C++ server is launched.
    \item The server is a wrapper for the model inference runtime llama.cpp. 
    \item The server manages all model-related inference computations.
    \item gRPC~\cite{grpc} protocol is used to establish the connection between the plugin and the server.
\end{itemize}

In mid-2023, we compared (1) ONNX Runtime with GPT-2-like models and (2) llama.cpp with LLaMA-like models, both with our native server implementation on MacOS with an M2 Max processor.
The benchmark showed that the mean time from the invocation of completion to the showing of the suggestion with ONNX Runtime was approximately 75 ms. 
For llama.cpp, the same procedure took approximately 100 ms.
Meanwhile, llama.cpp continued to be developed, and the same comparison dramatically changed in the late 2023: it became possible to infer the same LLaMA model in just 50 ms instead of 100 ms, so our development focus changed accordingly.

\insight{It is crucial to create a generalized infrastructure and regularly benchmark quickly developing frameworks to achieve the best possible product quality.}

The last unique feature of our implementation of inference is the \textit{caching of the model's hidden state}.
The beam search algorithm that we use consists of two steps: context processing and iterative generation.
Context processing, which computes token interrelations, is observed to be 3--10 times slower than generation (influenced by hardware and numerical precision).

To optimize the algorithm, we initialize the model at 50\% maximum context capacity, enhancing the speed of generation in two cases:
\begin{itemize}
    \item When the caret is moved within the initially processed context, generation proceeds with existing hidden states, conserving computational resources despite a potentially reduced context.
    \item Continuous code writing beyond the initial position leverages existing context computations, updating hidden states only for new code without extra processing.
\end{itemize}

To match the current context with the cached one, we maintain a list of integers corresponding to cached tokens.

This approach significantly improves efficiency by reducing redundant computations in the generation pipeline.
According to our observations, this allows to significantly increase model inference speed in over 90\% of the real-world scenarios, while keeping the quality at almost the same level.

\insight{Sophisticated caching strategies allow to speed up code completion without significant quality loss.}

\subsubsection{Post-processing}
\label{sec:approach:plugin:post}
In general, one can formulate the following two tasks for Full Line Code Completion:
\begin{itemize}
    \item Generate code suggestions so that the developer writes code easier.
    \item Do not disturb the developer with incorrect or unnecessary code suggestions.
\end{itemize}

Post-processing mostly targets the second goal: after the initial set of hypotheses is generated, we should decide whether it contains a useful suggestion, and if it does, which one it is.

\textbf{Filtering.}
\label{sec:approach:plugin:inference:filtering}
Our filtering mechanisms include:
\begin{itemize}
    \item \textbf{Safety}. Removes dangerous code (\textit{e.g.}, \textit{rm -rf, DROP DATABASE}), secrets (emails, access keys, etc.~\cite{secrets_entropy}), and profanity.
    \item \textbf{Low scored suggestions}. Eliminates suggestions that have a beam search score lower than a certain value.
    \item \textbf{Incorrect code}. Filters out unresolved references, type mismatches, incorrect argument counts, etc.
\end{itemize}

The \textit{Incorrect code} filter is the slowest one since it involves running code inspections, and thus it is launched the last and is restricted with maximum execution time.
Suggestions with undefined correctness status are not shown to the user.
Our research for the 2023.3 PyCharm Pro release showed that FLCC benefits from filtering out incorrect code. 
Only 1\% of valuable suggestions are lost, and as a trade-off for such a negligible loss, we establish the distinguishing feature of Full Line Code Completion --- \textit{suggesting only valid code}.

\textbf{Transformations.}
\label{sec:approach:plugin:inference:transformations}
After redundant suggestions are filtered out, we are left with a set of ``surviving'' hypotheses, among which we pick the one with the highest beam search score.
The chosen suggestion might still need to be tweaked a little bit, \textit{e.g.}, if not all paired symbols like brackets and quotes are correctly closed.
For this reason, in a final step before showing the code suggestion, we fix such incomplete code closures.

\subsection{User Experience}
\label{sec:approach:ux}
As mentioned before, seamless integration within the IDE is a primary objective for Full Line Code Completion. 
In pursuit of this UX goal, we have integrated inline completion functionality directly into the IntelliJ Platform API.
This allows us to create a proper integration with the standard IntelliJ completion, JetBrains AI Assistant~\cite{aia}, and third-party code completion providers. 
By the fall of 2024, Salesforce, Meta, and GitLab have already started using our API, and we continue negotiating it with other providers.
Using the same API allows to avoid visual conflicts between providers and to implement chain calls: if one provider cannot suggest code in the given position, another one tries to do so.

Our gray text functionality does not hide the standard code completion lookup window but reserves the \textit{TAB} hotkey for accepting code completion by default (see Figure~\ref{fig:completion}). 
Since this provokes usage conflicts, we implemented an option to change this shortcut.

Our plugin also shares some of the features with standard code completion, like auto-import, as described in Section~\ref{sec:approach:preprocessing}, and incorrect code detection, as described in Section~\ref{sec:approach:plugin:inference:filtering}.

Last but not least, our implementation of inline completion  supports several hacks and tricks to ensure a smooth user experience. 
This includes: \textbf{(1)} \textit{seamless overtyping}, \textit{i.e.}, making sure the suggested code does not blink when you type; \textbf{(2)} \textit{computation delay}, so the developer does not accept a sudden code suggestion with the \textit{TAB} keystroke while intending to increase indentation; and others.

Overall, our approach to implementing Full Line Code Completion allows us to cover all restrictions and design principles described at the beginning of this section, meaning that the resulting feature is local-based, fast, compact, and properly integrated into the IDE.

%% file: sections/04-evaluation.tex
\section{Evaluation}
\label{sec:evaluation}
In order to iteratively improve Full Line Code Completion, we established several feedback loops, including \textit{offline} and \textit{online} evaluation.
Offline evaluation utilizes only a dataset with source code and is frequently used in research because the academic community usually does not have access to a large user base. 
Online evaluation involves collecting data from real users and is more likely to be used for products.
In FLCC, we first check the pipeline quality using offline evaluation, however, we pay a lot more attention to online evaluation. 
For this reason, in this section, we briefly describe the setup of offline evaluation and give one example of how we used it to evaluate a strategy of collecting hypotheses, and then focus more on the online evaluation. 
It is important to note, however, that we cannot describe all the details of evaluating different models and setups within the scope of this paper, so this section serves to convey the most important results and the overall pipeline of how such features are evaluated in industry.

\subsection{Offline Evaluation}
\label{sec:evaluation:offline}
When training our model, we calculate recall and mean reciprocal rank, as described in Section~\ref{sec:approach:model_training}.
However, these metrics are single-token and evaluate only the quality of the model, rather than that of the entire code completion pipeline, which is what is actually needed to test the feature.

For offline evaluation, we use several repositories from the test segment of the collected dataset. 
In them, we pick a list of files of interest, and in each file, we generate various caret positions, where code completion is automatically invoked and evaluated.
The picked repositories, files, and caret positions are fixed for every evaluation launch, so it is possible to track changes and study the model's performance from iteration to iteration. 
We use an automatic IDE launch and programmatically invoke code completion.
This enables us to use the entire pipeline implemented in the FLCC plugin, including caching, filtering, and other features.
Moreover, this allows not only to compare different versions of the plugin, but also to compare FLCC with standard code completion.

Metrics that we track with such a setup form a long list, which includes the following most important ones:
\begin{itemize}
    \item \textbf{Matched ratio.} Ratio of symbols matched with the actual code to the right of the position where it was invoked.
    \item \textbf{Perfect lines.} Ratio of FLCC invocations that fully match the actual code to the right of the invocation position.
\end{itemize}

With this setup, we run experiments for every major change implemented in FLCC.
For example, while implementing our strategy for collecting hypotheses described in Section~\ref{sec:approach:sequence:stashing}, we observed that, for Python, it increased the percentage of Perfect Lines from 13\% to 21\%, while Matched Ratio only dropped from 20.6\% to 20.3\%, which is quite negligible.

\subsection{Online Evaluation}
\label{subsec:evaluation:online}
Offline evaluation is a valuable and convenient tool, but it might be biased, so it is not enough to make reliable decisions about product changes. 
As described in Section~\ref{sec:bg_and_rw:bg}, at JetBrains, we use early access programs (EAPs) to collect anonymous logs for experimental features and evaluate their performance.
For ML features, in particular, we pay attention to rigorous A/B testing during EAPs.
Such an approach allows to compare user experience with only standard completion versus both standard and Full Line Code Completion enabled. 
Also, this provides a convenient way of comparing different implementations of FLCC itself. 
During such A/B testing in late 2023 and 2024, we collected a lot of data to compute different important metrics from literature~\cite{ziegler2022productivity, wang2023practitioners, codeium-cpo}.
We share the overall comparison for multiple languages in Table~\ref{tab:evaluation} and describe the metrics below, highlighting results for Python.
All the provided comparisons are statistically significant, measured by employing per-user bootstrap~\cite{efron1992bootstrap}.

We would like to emphasize that the choice of the proper online metric is a difficult task. 
While choosing the metrics, we paid attention to two goals. 
Firstly, metrics should be easy to compute and well-adopted in the community to put FLCC on the scale with other available code completion solutions. To that end, we employed the acceptance rate (AR).
Secondly, we wanted to incorporate metrics that are agnostic to the source of code completion or a particular UX implementation (\textit{e.g.}, lookup completion or gray text completion).
Thus, we used the ratio of completed code (RoCC) to be able to compare IDE code completion with and without FLCC.

\begin{table}[h]
\caption{The results of online evaluation for different languages.}
\centering
\begin{tabular}{cccc}
\toprule
\multirow{2}{*}{\textbf{Language}}& \multicolumn{2}{c}{\textbf{Ratio of completed code}} & \textbf{Acceptance rate} \\ \cmidrule(lr){2-3} \cmidrule(lr){4-4}
	& \textbf{without FLCC} & \textbf{with FLCC} & \textbf{with FLCC}  \\ \midrule
Java & 0.33 & 0.38 $\color{OliveGreen} \uparrow$	& 0.32 \\
Kotlin	& 0.28 & 0.30 $\color{OliveGreen} \uparrow$ & 0.23 \\
Python & 0.17 &  0.22 $\color{OliveGreen} \uparrow$ & 0.38 \\
JS/TS/CSS & 0.21 & 0.27 $\color{OliveGreen} \uparrow$	& 0.24 \\
C/C++ & 0.16	& 0.22 $\color{OliveGreen} \uparrow$ & 0.18 \\
Go & 0.23	& 0.29 $\color{OliveGreen} \uparrow$ & 0.38 \\
PHP	& 0.20 & 0.27 $\color{OliveGreen} \uparrow$ & 0.31 \\
Rust & 0.20 &  0.27 $\color{OliveGreen} \uparrow$	 & 0.28 \\
\bottomrule
\end{tabular}
\label{tab:evaluation}
\end{table}

\textbf{Ratio of completed code.} 
This is our main, golden star metric used for the assessment of code completion quality. 
It is defined as a ratio of symbols of code written with code completion among all the written code.
The numerator of the metric includes code from all code completion sources: standard and inline completions, so it allows us to assess the entire code completion experience.

For Python, the metric bumped 1.3 times (from 17\% to 22\%) for users with FLCC enabled compared to users with standard code completion only.

\textbf{Acceptance rate.} 
This is the ratio of events where Full Line Code Completion was selected to events where it was shown.
Previous studies~\cite{ziegler2022productivity} show that the acceptance rate strongly correlates with the perceived quality of code completion.

The acceptance rate for Full Line Code Completion for Python was about 38\%.
For standard completion, the value is about 27\%, however, no conclusions can be made from this comparison due to significantly different lengths of suggestions. 
For comparison, GitHub Copilot shows an acceptance rate of about 30\%~\cite{copilot-ar} for both single-line and multi-line suggestions combined.
However, one should be careful when comparing LLM-based completions using AR only, because this metric does not account for suggestion length or performance.

\textbf{Persistence rate.} 
This metric can be defined differently~\cite{ziegler2022productivity}, 
and it accounts for the ratio of code that stays in the file for some time after accepting code completion.
In our approach, we track the average number of cancellation actions, including Ctrl/Cmd+Z shortcuts, and various text deletion events after the code completion suggestion was selected.

We did not observe any statistically significant data that would show that users cancel completion selections or delete text more frequently when they are provided with FLCC.

\textbf{Time to show.} 
This metric calculates CPU time from the invocation of completion to the showing of a suggestion.

The average time to show for FLCC is 150 ms, which is perfect for the user's visual perception~\cite{kosinski2008literature}. 
The result includes cache initialization events, as well as cache reusing. 
According to our observation, the cache is used in over 90\% of completion events, which makes a typical time to show even lower.

\textbf{IDE performance.} 
Using our internal quality gates, we tracked the performance of various IDE events, such as standard code completion and the IDE launch.
We did not observe any performance degradation when enabling FLCC.

%% file: sections/05-discussion.tex
\section{Discussion}
\label{sec:discussion}

\textbf{Impact of Full Line Code Completion.}
The enhanced efficiency introduced by FLCC in software development is the key finding of our study.
The feature, as evidenced by our A/B tests, proves more effective than the standard single-token approach, offering faster coding and potentially reduced cognitive load for developers. 
However, it is important to recognize that the integration of such advanced tools also requires developers to adapt to new workflows, which might initially pose a challenge for some users.

Also, the scope of the implemented feature is limited to a single line of code.
On the one hand, this constraints the possible impact of the feature, because it cannot generate bigger code snippets.
On the other hand, the given limitation makes the feature controllable, meaning the relevance of the suggested line can be easily assessed by a developer.
Additionally, by providing such a feature as a built-in functionality, we might raise the further adoption of AI-driven tools in the community.

\textbf{Challenges and future directions.}
Despite the evident advantages, implementing FLCC in a local environment presented unique challenges, especially considering the constraints of memory footprint and processing speed. 
While we overcame these challenges, they highlight the need for optimizing ML models for practical use in constrained environments.
Additionally, the feature's performance is significantly different for different languages, so we are aiming to study these differences and gain more insights about them.

\textbf{Bridging the gap between research and practice.}
Our work with FLCC also provides valuable lessons about the integration of complex, research-based solutions into commercial products. 
Our experience can serve as a guide for both academia and industry, highlighting a gap that often exists between theoretical model performance and practical deployment. 
Furthermore, the developed gray text API and our approach to evaluating FLCC based on users' data could be a model for future endeavors in the field, ensuring that such advanced tools not only pass the technical benchmarks but also align with user needs.

%% file: sections/06-threats.tex
\section{Threats to Validity}
\label{sec:ttv}
The industrial context of our task and the production-focused nature of our solution place certain constraints on our work. 

\textbf{Limitations of offline evaluation.}
We use a limited number of repositories and files to run our offline evaluation pipeline. 
Despite our effort to collect repositories with diverse code, our dataset for offline evaluation remains limited in terms of size and might not represent all the code well enough.

\textbf{Online evaluation bias.}
Our study utilized anonymous data from Early Access Program (EAP) users to evaluate various code completion approaches via A/B testing. 
Although the results were statistically significant, they may not fully represent the broader user base of IntelliJ-based IDEs, as EAP participants typically have a deeper understanding and engagement with the IDE's features, and their usage patterns may differ from average users.

\textbf{State of the art.}
In contrast to most approaches in academic research, our focus extends beyond merely training a neural network for the code completion task.
We aimed to develop the code completion \textit{feature}, taking into consideration the behavioral patterns and preferences of users interacting with an IDE and its built-in code completion functionality.
It is important to note that our analysis is limited to a specific group of users --- those participating in the EAP.
This limitation restricts our ability to conduct a direct comparison with other multi-token completion tools, which are used by a broader user base not involved in the EAP.

%% file: sections/07-conclusion.tex
\section{Conclusion}
\label{sec:conclusion}
In this work, we shared our approach to implementing a multi-token code completion feature for IntelliJ-based IDEs, called Full Line Code Completion. 
The feature fills a niche of local-based multi-token code completion engines, being available for a wide range of users, including those with limited internet access and those with privacy concerns. 
It meets the speed and memory footprint restrictions, as well as fulfills several other important design principles.
The resulting implementation has proven its efficiency during A/B testing on real users and was bundled into 2024.1 \& 2024.2 releases of IntelliJ-based IDEs. 
Our ongoing and further work is focused on different languages and making sure that the feature performs well in all of them.
We hope that our work can prove useful to both researchers and practitioners, and showcase the lifecycle of a complex research project in the real product.